# Streamlined Calibrations of the ATLAS Precision Muon Chambers for Initial LHC Running

N. Amram[1], R. Ball[2], Y. Benhammou[1], M. Ben Moshe[1], T. Dai[2], E.B. Diehl[2], J. Dubbert[3],
E. Etzion[1], C. Ferretti[2], J. Gregory[2], S. Haider[4], J. Hindes[2], D.S. Levin[2], R. Thun[2],
A. Wilson[2], C. Weaverdyck[2], Y. Wu[2], H. Yang[2], B. Zhou[2], S. Zimmermann[5]

1. Raymond and Beverly Sackler School of Physics and Astronomy, Tel-Aviv University, 69978 Tel Aviv, Israel

2. Department of Physics, The University of Michigan, Ann Arbor, MI 48109-1120, USA

3. Max-Planck-Institut feur Physik, (Werner-Heisenberg-Institut), Muenchen, Germany

4. CERN, CH - 1211 Geneva 23, Switzerland

5. Fakultat fuer Mathematik und Physik, Albert-Ludwigs-Universitat, Freiburg, Germany

*January 2011*

***Abstract -*** *The ATLAS Muon Spectrometer is designed to measure the momentum of muons with a resolution of dp/p = 3% and 10% at 100 GeV and 1 TeV momentum respectively. For this task, the spectrometer employs 355,000 Monitored Drift Tubes (MDTs) arrayed in 1200 Chambers. Calibration (RT) functions convert drift time measurements into tube-centered impact parameters for track segment reconstruction. RT functions depend on MDT environmental parameters and so must be appropriately calibrated for local chamber conditions. We report on the creation and application of a gas monitor system based calibration program for muon track reconstruction in the LHC startup phase.*

## 1  Introduction

The design momentum resolution of the ATLAS Muon Spectrometer [1, 2], dp/p=3% and 10% at 100 GeV and 1 TeV momentum respectively, is attained through the combined efforts of a sophisticated optical alignment system by dedicated calibration centers tasked with providing daily updated drift tube time-to-radius (RT) functions and timing offsets [3]. The latter are required by the precision Monitored Drift Tube (MDT) tracking chambers to achieve the intrinsic 80 micron tube hit resolution. The calibration centers, located in Europe and the U.S., consist of large computing farms feeding from high rate data streams derived from the secondary ATLAS muon trigger. In full LHC operation these centers are intended to provide optimal MDT calibrations. For early LHC running where luminosity and the associated muon production are lower than the design specifications, we have developed a streamlined calibration program that augments the functioning of the calibration centers. This program is based on a dedicated MDT gas monitor chamber, whose operational details are reported elsewhere [4]. For this purpose, the gas monitor chamber provides daily function calibration constants or functions that relate the measured drift time to the drift radius for a particle track crossing a tube. The functions are generally referred to as RT functions. After drift time corrections for local temperature, pressure and magnetic field, these RTs are applied to all MDT chambers. These chamber-specific RT functions characterize the MDT gas for a given 24-hour period. The local drift time corrections are determined from daily averaged chamber temperatures and from magnetic field vectors, which are determined from on board sensors and field maps. Correction functions are determined from GARFIELD [5] / MAGBOLTS [6] based calculations, which have been validated with data. Various tests, including track quality assessment from residual distributions of cosmic ray runs, have shown this program to provide very good calibrations suitable for the startup phase of the LHC when luminosities are low. The remainder of this article, after a brief description of the Muon Spectrometer MDT chambers and gas monitoring system, details the components of the streamlined calibration program.

## 1.1 ATLAS Muon Spectrometer

The ATLAS Muon Spectrometer is a cylindrical detector, 45 m long and 22 m in diameter. It is divided into three regions, the barrel and two endcaps [2]. Muons are deflected by a magnetic field produced by three large superconducting air-core toroids; each toroid (two endcap magnets and one barrel magnet) consists of eight coils. The strong bending power of the magnetic system reaches ~5.5 Tm in the barrel region. The Monitored Drift Tube chambers provide the primary track coordinate measurements. An MDT chamber consists of two multilayer of three or four tube layers of densely packed drift tubes. The 1200 MDT chambers comprise nearly 355,000 individual drift tubes: cylindrical, one to six meter aluminum tubes filled with 3 bar Ar 93%, $CO_2$ 7% gas with a 3080 V high voltage wire at the center, and operated at a gain of 20,000. For triggering and measurement of the second coordinate, Resistive Plate Chambers (RPCs) are used in the barrel and Thin Gap Chamber (TGCs) are used in the endcaps [2]. The barrel comprises three nested, cylindrical sections in which chambers are oriented around the beam axis (z-direction) with MDT wires along the $\phi$ (azimuthal) coordinate. The endcaps are divided into three wheels with MDT wires along the phi coordinate, but with chambers oriented perpendicular to the beam axis.

## 2 Gas Monitor Chamber Calibration Program

The gas monitor chamber performs two tasks. First, it continuously analyzes the MDT gas drift spectra and provides hourly updates of gas quality [4, 7]. Secondly, it generates a URT function every two hours corresponding to the standard temperature and pressure of 20 $C^0$ and 3 bar. The URT function represents a calibration baseline for the MDT gas at any time [8]. From each day's 12 URTs, an average URT is constructed. From the averaged URT, compensating the average URT for each chamber's average temperature, pressure, and magnetic field makes a chamber-level RT function. These chamber-level calibrations can be used for track reconstruction. The gas monitor chamber calibration procedure is a byproduct of the normal gas monitoring tasks, and so requires minimal additional computational resources.

## 2.1 Drift Times to Track Segments

The spectrometer's precision coordinate is transverse to the chamber plane and is determined by the radial distance to the anode wire of a charged particle's track, which passes through an MDT. Ionization electrons accelerate towards the wire, initiating a charge avalanche, and producing a signal.

The drift time is defined as the time of arrival at the wire of the first drift electrons along a trajectory corresponding to the distance of closest approach to the passing ionizing particle. This drift time is the primary measured physical quantity in the MDT system. The read-out time is the threshold crossing time of the anode signal. The minimum drift time is this threshold crossing time minus that of a zero impact parameter track, one passing at the
tube center. The latter is called a T0 and represents a timing offset. The T0 is a function of cable delays and particle times of flight.

The drift radius is computed from the drift time by a time to radius conversion function, RT function, which is computed through an auto-calibration procedure. To assign a track segment, the drift radii are connected from each hit MDT by a line, which passes tangent to each drift circle. The track is defined as the line which minimizes the track residuals, track radius minus drift radius, and is computed in a calibration procedure iteratively. A typical track segment through a chamber can be seen in Figure 1.

To achieve optimal resolution over the entire Spectrometer, the RT functions must be tuned to the local track segment conditions. Ideally the RT functions should be determined for a region over which these conditions are homogeneous. In practice these regions generally correspond to a single chamber or chamber multilayer. In ATLAS, the local chamber environmental conditions are measured from embedded sensors, and are accounted for when building an RT function calibration program.

## 2.2 Gas Monitor System

The first purpose of the gas monitor chamber is to track the temporal change in the MDT system gas, which causes significant drift time changes. It will be shown that other causes of drift time variations can be accurately addressed. Therefore the primary, long term deviation of compensated drift times is due to gas composition alteration. Composition change is primarily water injection and $CO_2$ concentration fluctuation [7]. Other variation, such as the influx of atmospheric diatomic molecules $O_2$ and $N_2$, are harder to model, but diffuse at a significantly slower rate and so represent much smaller deviations. Diffusion of water vapor on the order of 1000 ppm can increase drift times by as much as 10%. Similarly, $CO_2$ concentration increase of 0.5% can increase drift times by nearly 10%. As a result gas composition changes, if not carefully tracked, can cause significant MDT resolution degradation; this makes monitoring MDT gas essential. Moreover, an ideal drift time correction algorithm would be based upon RT functions that represent the spectrometer's gas composition at any given time, whatever it may be. The gas monitor system provides such RT functions.

### 2.2.1 Gas Monitor Chamber

The gas monitor chamber has the same essential features of any active chamber in the spectrometer, though on a smaller scale. The chamber and its drift tubes were constructed at the University of Michigan in the same way as the ATLAS endcap chambers. 96 drift tubes were glued into a pair of three-layer multilayers, with dimensions of (50, 70, 21) cm. This mini-chamber is further partitioned into two identical units, each with separate connections to the gas system: one to the input gas line and the other to the return line. Temperature and pressure changes are measured with 0.1 K and 1 mbar precision respectively from several temperature sensors, and pressure sensors placed at the output and input gas ports. A depiction of the gas monitor chamber can be seen in Figure 2. The Muon Spectrometer gas monitoring is achieved by strategic placement of the gas monitor chamber in the ATLAS gas facility, a surface building located 100 m above and 50 m displaced from the experimental cavern. The input and return line gas in the two partitions represents an accurate sample from all MDT chambers because the gas system supplies and exhausts all MDT chambers in parallel. The difference in gas between the two lines has been found to be small and temporally stable [7]. The gas monitor chamber acquires cosmic ray muon data from a scintillator trigger, two plastic scintillator panels positioned above and below the chambers each connected to a phototube, at 15 Hz. Every hour some 50,000 tracks are collected, during which time the gas volume has been mostly renewed [4].

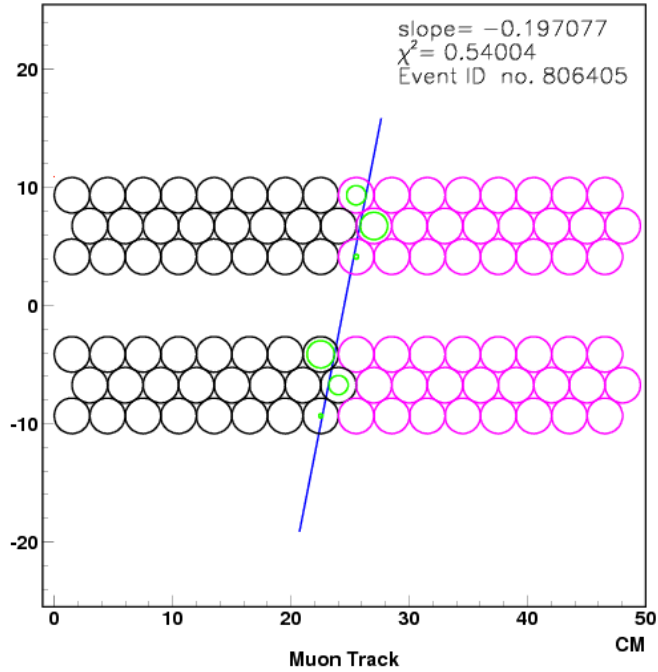

**Figure 1: Typical chamber track reconstruction. Track passes tangentially through each drift circle**

### 2.2.2 MDT Parameter Analysis with $T_{max}^{spectrum}$

The drift time spectrum is sensitive to the MDT gas composition and can be analyzed as such provided certain assumptions about the control and stability of the gas components. The spectrum is constructed from the muon cosmic ray events in the gas monitor chamber. The spectrum used is an aggregate from all 48 drift tubes in each partition. The hourly spectrum for each partition has some 150,000 entries. A typical gas monitor drift time spectrum can be seen in Figure 3. From this spectrum a set of parameters are extracted which characterize the chamber conditions. The most pertinent to the MDT environmental parameter analysis is the maximum drift time, $T_{max}^{spectrum}$. This parameter is calculated by fitting the rising and falling edges of the spectrum with a modified Fermi-Dirac function (Eq. 1).

$$f(t) = \frac{A+Dt}{1+e^{\frac{B-t}{C}}} \quad (1)$$

The coefficient B represents the 50% rise/fall time, coefficient C represents the width of the rise/fall time, and Coefficient D represents a slope before the tail of the distribution. The difference between the B coefficients from the rising and falling edge fits is defined as $T_{max}^{spectrum}$. The typical statistical error in the measurement of $T_{max}^{spectrum}$ is ~0.6ns, or 0.1%. The $T_{max}^{spectrum}$ represents the average electron drift velocity across the tube, and thus, after the effects of gas monitor chamber temperature and pressure are accounted for, it can be used as a sensitive measure of gas composition. For example, the addition of 100 ppm of water vapor increases $T_{max}^{spectrum}$ by 7 ns, or ~1%, which is nearly an order of magnitude larger than a typical, random statistical error.

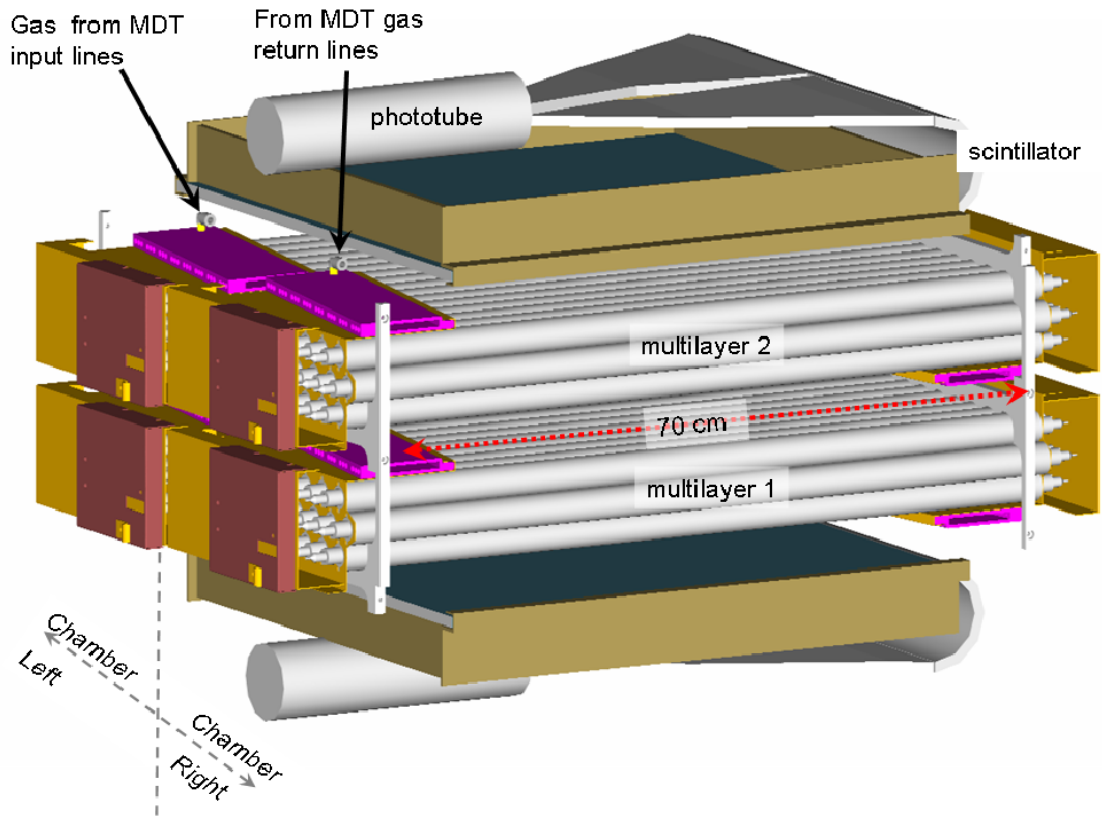

**Figure 2: Drawing of Gas Monitor Chamber.**

The ATLAS MDT installation took place until July 2008. The time span November 2008 to March 2009 was spent for commissioning water injection procedure to the MDT system. A new and improved water injection system was installed in November 2009. The MDT system was running in stable conditions since January 2010. Traces of these changes were recorded by the gas monitor system. The gas monitor chamber has been in nearly continuous operation since September 2007. For more than three years, a reliable image of the MDT gas system performance has been established. Figure 4 shows the temporal variation in $T_{max}^{spectrum}$ for over a three year period, including the first year of LHC running. It illustrates the gas composition stability and the small difference (0.1% of $T_{max}^{spectrum}$) between the return and supply lines. The $T_{max}^{spectrum}$ is observed to vary on all time scales: days, weeks, months, and years. The variations are due primarily to the change in water vapor in the gas mixture. The source of the water is from ambient humidity and intentional water injection to nominal level of 1000 ppm [9]. The water injection is needed to prevent cracks in the plastic endcaps of the drift tubes and to avoid local built-up of charges. Drift time changes roughly follow changes in the cavern's humidity. Typically daily variation is ~1 ns, while weekly variation is ~10 ns [7].

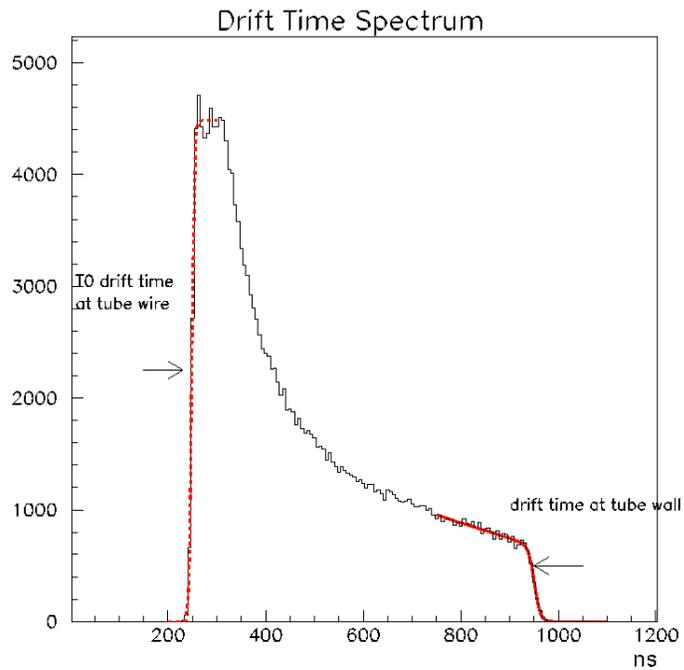

**Figure 3: GMC drift time spectrum with Fremi Dirac function fit.**

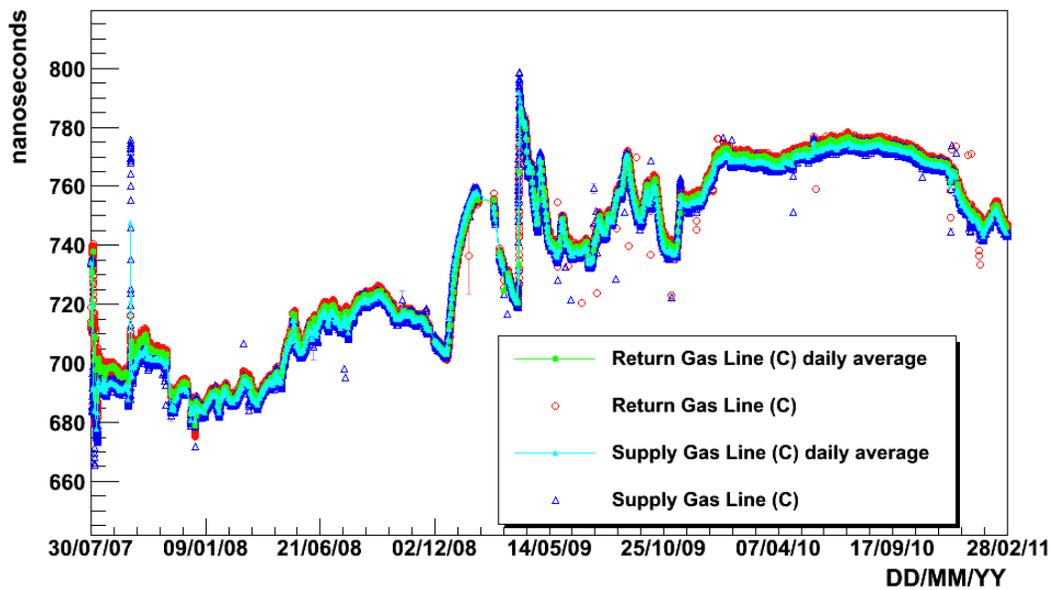

**Figure 4: Temporal variation of GMC $T_{max}^{spectrum}$ with return and supply lines from 18/03/2007 to 28/02/11. This range includes the first year of LHC beam collisions from March 30, 2010.**

### 2.2.3 Gas Monitor RT Function Generation

RT functions represent a mapping from drift times, which are directly measured in the Muon Spectrometer, to drift radii through the electron drift velocity. Gas monitor RT functions are generated from an auto-calibration procedure in the
analysis program mutrak. Every two hours some 100,000 tracks from cosmic ray muons are fit. The procedure tries to find the best track fit for all the tracks. The average track fit residuals from the collection of events are determined at each of 100 radial bins spanning from the tube wire to the tube wall. Track fitting requires an initial estimate to the RT function. The initial RT function can be one determined under different gas conditions or can be derived directly from the integral of the drift spectrum dN/dT, and assuming a uniform flux, dN/dR=constant: dR/dT = dR/dN × dN/dT. In practice, the initial RT is simply the previously computed RT from the most recent dataset. The starter RT is changed by adding the track residuals at each bin iteratively until the optimal RT is produced. Convergence is assumed when there is no change in the maximum drift time of the RT function, $T_{max}^{RT} = T(tube\ wall\ radius \sim 14.6\ mm)$ upon further iterations. An example of an RT from the gas monitor chamber is shown in Figure 5.

## 2.3 GARFIELD Simulations and Normalized Correction Functions

GARFIELD is a computer program, which can simulate gaseous ionization detectors such as the ATLAS MDTs. It was used to model the muon system drift tubes with varying operation parameters: gas composition, wire sag, high voltage, magnetic field, temperature, and pressure. The variation analysis was carried out using simulated RT functions. The operation parameters were varied in a controlled way over the possible operation ranges. Statistical fluctuations in the RT function are inherent from Monte Carlo integration and random number generators used by GARFIELD. Consequently, a spread in $T_{max}^{RT}$ on the order of nanoseconds is not uncommon for several simulations with the same conditions. Operation parameter alterations typically cause drift times to vary on the same scale, so a method was needed to flatten the inherent statistical error.

### 2.3.1 General Simulation Procedure

In order to correct RT functions for a particular field variation, we require a calibrated correction function that can be added, bin-by-bin, to a standard condition RT function. The drift times are corrected as a function of radius because the range of the drift radius is fixed, while the range of the drift time is always changing. The correction function is built from an ensemble of GARFIELD simulated RT functions. To minimize statistical error the MAGBOLTS interface collisions parameter in GARFIELD is set to 100, from its default 10. The random error is minimized by a factor of $\sqrt{10}$, but the run time grows linearly. An ensemble of RT functions is generated for each set of field conditions from which an average RT function is constructed, by averaging the ensemble for every bin. Each set of 11 averaged-RT functions represents 11 different field conditions.
From the 11 distinct RT functions 55 unique difference functions, or dTR functions, can be computed. The dTR curves are computed by first inverting and then fitting each averaged RT function with a series of Chebyshev polynomials. The typical fit residuals to the RT functions are everywhere less ~0.1 ns, or an order of magnitude smaller than the minimum drift time corrections. Moreover, the residuals are fairly uniform across the tube. Plots of a typical RT function mapped into Sobolev / Chebyshev space, [-1,1], and fit residuals are given in Figure 6 and Figure 7. The 55 dTR curves are computed by taking bin-by-bin differences between all combinations of the 11 averaged, Chebyshev TR polynomials. The dTR curves are normalized through a procedure that depends on the type of field variation. However, the end result is one correction function that can be scaled to give the change in drift time for any radial bin and any field value.

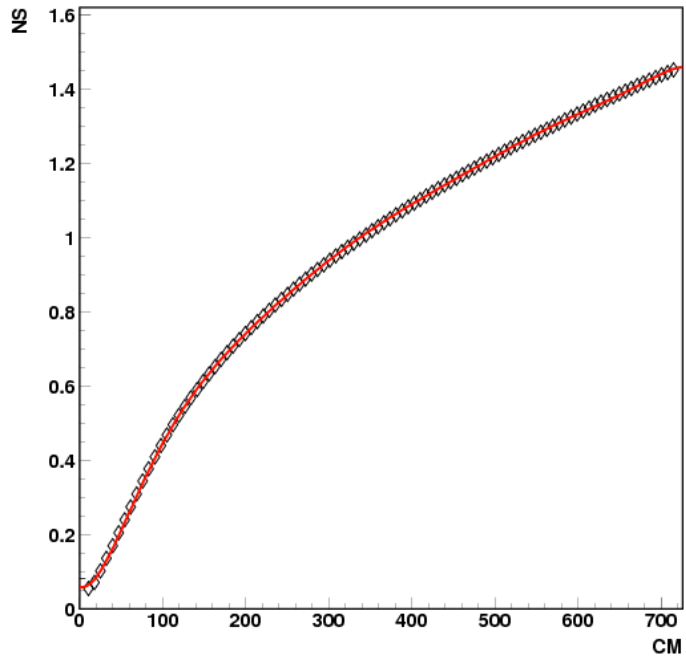

**Figure 5: Gas Monitor MDT RT function (3bar, 93/7 Ar $CO_2$ , 293K)**

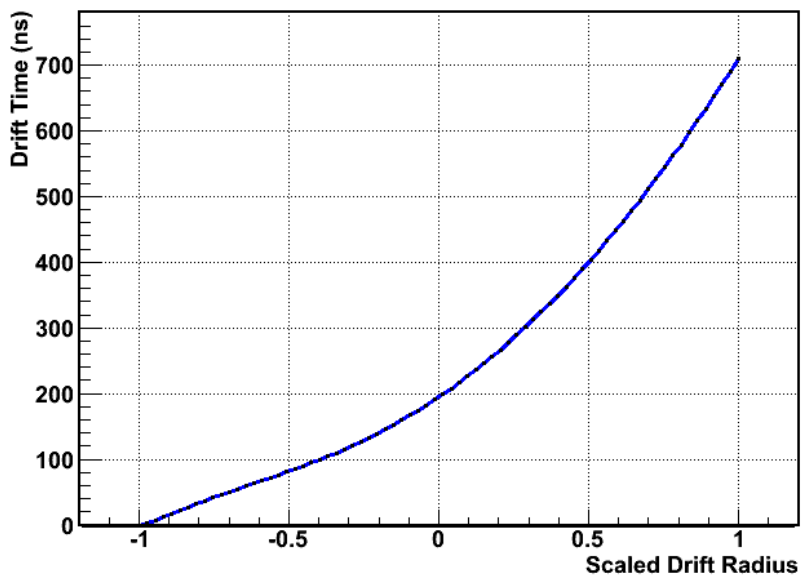

**Figure 6: GARFIELD generated RT-function mapped into Sobolev/Chebyshev space. The range (-1,1) corresponds to radius = 0 at the anode wire to 14.6 mm at the tube wall.**

### 2.3.2 Temperature and Pressure Corrections

The ATLAS Muon Spectrometer spans six stories with many heat sources and sinks. Consequently, chamber temperatures deviate from 293K. The dominant variance is an average vertical gradient of $(7/22)\frac{K}{m}$ [10]. Figure 8 shows the distribution of temperatures in the spectrometer. Each point represents an average measurement from several sensors. The MDT gas pressure is regulated to within a few mbar of 3bar. The electron drift velocity is proportional to $\sqrt{(E/p)}$, where E is the applied electric field and p is the gas pressure. Therefore, the fractional change in drift velocity is proportional to the fractional change in density, temperature, and pressure: $dv/v \propto -d\rho/\rho \propto dT/T \propto -dp/P$ [11]. Consequently, drift time changes of this sort are linear. Moreover, temperature and pressure changes are nearly equivalent because they manifest themselves as changes in gas density, or mean free path, though in opposite directions. An increase in pressure of 10 mbar is approximately equal to a decrease in temperature of ~1K. Fluctuations in both can vary drift times from 0.5% to 5%. For example, $dT_{max}$ decreases ~3 ns per 1.2K temperature increase.

#### 2.3.2.1 GARFIELD Temperature and Pressure Variation Simulations

The linearity of the changes in drift times with variations in pressure and temperature can be verified using GARFIELD / MAGBOLTS simulations. A plot of $dT_{max}^{RT}$ versus pressure change and $T_{max}^{RT}$ versus temperature from GARFIELD simulated RT functions are shown in Figure 9 and Figure 10 respectively. The change in $T_{max}^{RT}$ is 0.247 ns/mbar and -2.51 ns/K for pressure and temperature changes respectively. We follow the basic correction procedure explained above to construct pressure and temperature RT function corrections. The changes in drift times are radius dependent, but each correction scales by the same factor that the temperature and pressure changes scale. In this way, once a bin-by-bin calibrated correction function is constructed for a particular temperature change, any arbitrary correction must only be scaled by a factor of $(\Delta T\_observed)/(\Delta T\_baseline)$ for an observed temperature change $\Delta T_{observed}$, or $\alpha * \Delta P_{observed}/\Delta T_{baseline}$ where $\alpha$ =0.0992 K/mbar for an observed pressure change of $\Delta P_{observed}$. Because the temperature and pressure are known to 0.1% and a few mbar respectively throughout the spectrometer for every MDT chamber, these corrections could be used to correct the baseline gas monitor RT functions for pressure and temperature variation.

The algorithm for computing temperature and pressure corrections proceeds as follows:
1. An ensemble of RTs are generated and averaged for each GARFIELD temperature simulation, 287 K to 298 K in 1.2 K steps
2. The RTs are mapped into Sobolev space and fit with Chebyshev polynomials (see II.3.A).
3. Iterative difference functions, dTR curves, are computed for every possible temperature change combination.
4. The dTR curves are normalized to the same base-line correction $dTR(\Delta T = 1.2\ K)$ by convention.
5. Each dTR bin has 55 points which are averaged, filtered with a 2*sigma cut, and then re-averaged.

The remaining calibration curve represents a scalable, radius-bin-by-bin, drift time correction function that can be added to the 293 K:3 bar gas monitor RT functions to give a drift time-drift radius calibration that is compensated for gas composition, temperature and pressure. Figure 11 shows a set of difference curves computed relative to 283 K, TR(283 K). The curves are equally spaced which signifies the linearity of the changes in drift times with temperature/pressure at all drift radii. The calibrated correction function can be seen in Figure 12 - Note that it has been normalized to an increase in temperature of 1.2 K.

An internal consistency check of our algorithm with GARFIELD can be made by adding our properly normalized calibration function to a baseline RT, and measure how well it approximates a simulated RT function with a considerably larger temperature. Figure 13 shows the residuals between a simulated RT function at 299 K, RT(299), and an RT formed by the addition of an RT function simulated at 287 K, RT(287), plus the calibration function normalized to a change in temperature of 12K. Comparatively, this is a large temperature difference, $T_{max}^{RT}$ is ~30 ns. The residuals shown in Figure 13, are very small across the tube and at most ~1 ns near the tube wall. The induced error is at most 5% of the correction and 0.1% of the maximum drift time. However, smaller temperature differences will have smaller drift time disparities. In addition, a direct test of the correction function was performed by comparing it with the difference between two Gas Monitor RT functions with a known temperature difference, -3.1 K. Figure 14 shows the very good agreement between the difference curve and the correction curve. The disparities are mostly within the error and likely due to the uncertainty in the temperature measurements and the analysis program's pathologies near the tube wall.

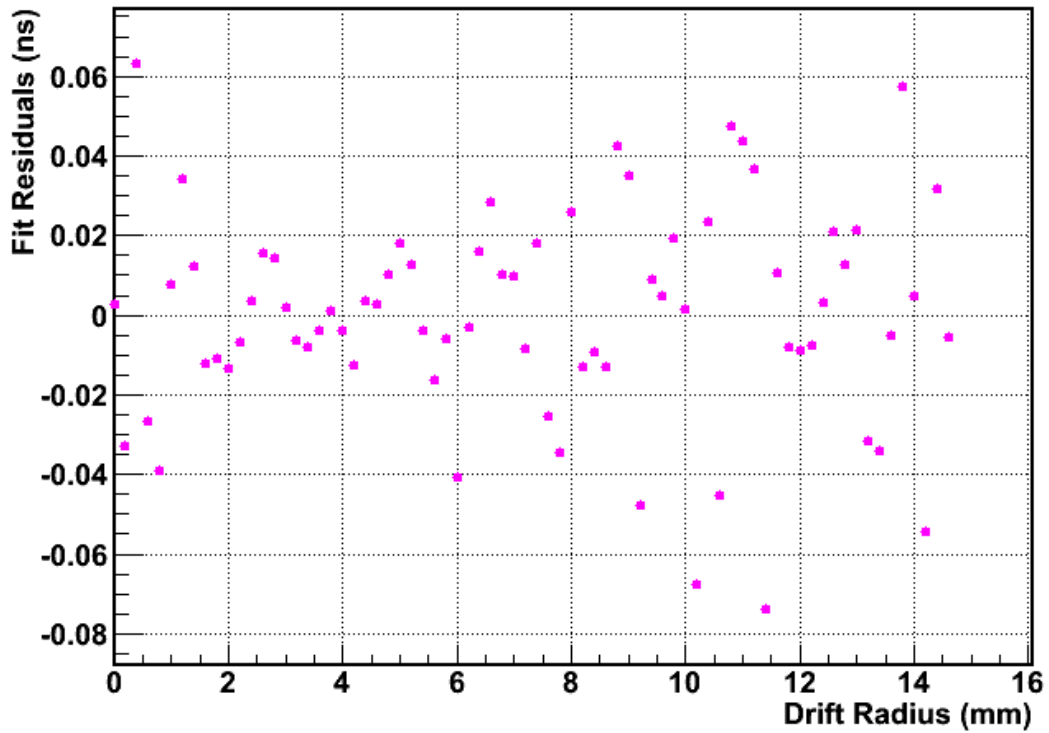

**Figure 7: Chebyshev polynomial fit residuals to GARFIELD generated RT-function versus drift radius.**

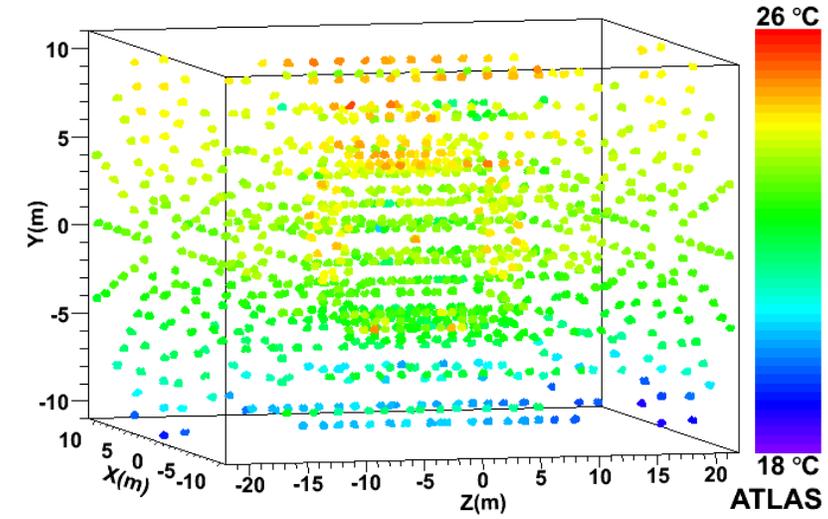

**Figure 8: Temperature Distribution in the ATLAS Muon Spectrometer.**

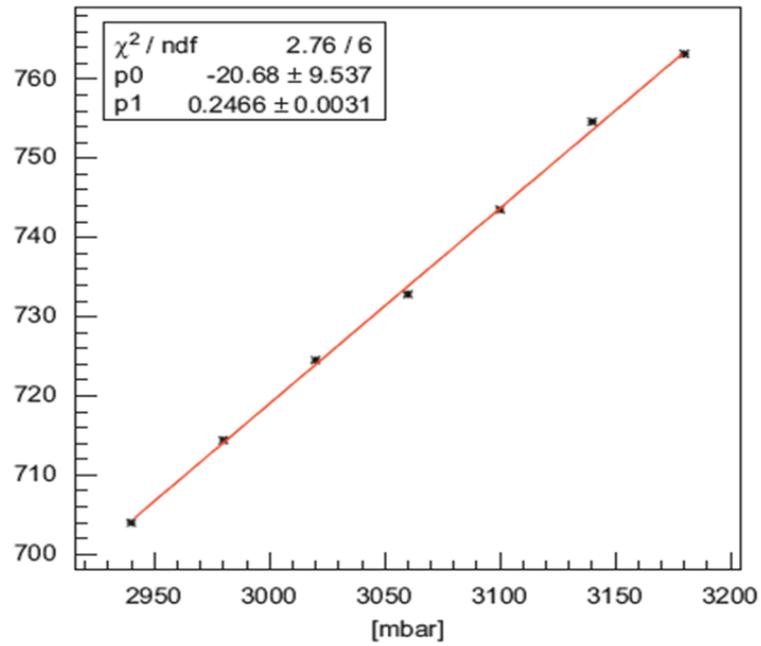

**Figure 9: Maximum drift time as determined from the endpoint of the GARFIELD generated RT-function versus MDT pressure.**

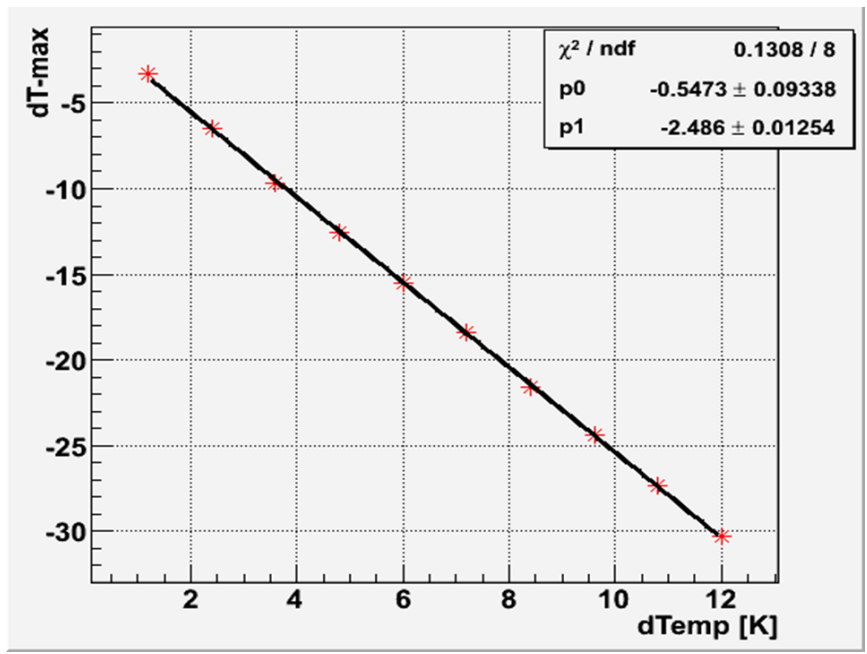

**Figure 10: Change in maximum drift time of GARFIELD generated RT-functions from (3bar, 93/7Ar $CO_2$, 293 K).**

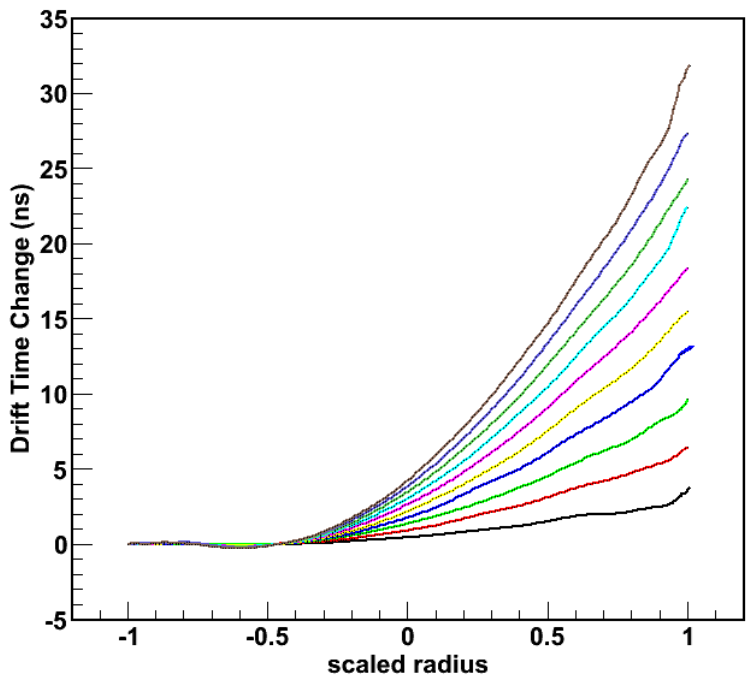

**Figure 11: Drift time differences between GARFIELD RT functions versus drift radius: TR(287)-TR(287+n*1.2 K) where n is an integer from 1 to 10.**

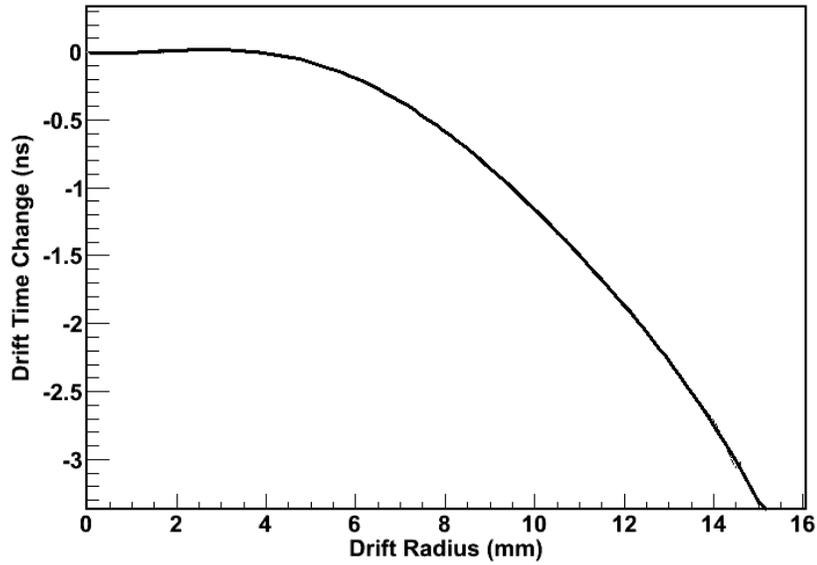

**Figure 12: Temperature Calibrated drift time correction function from GARFIELD generated RT-functions normalized to a 1.2 K MDT temperature increase.**

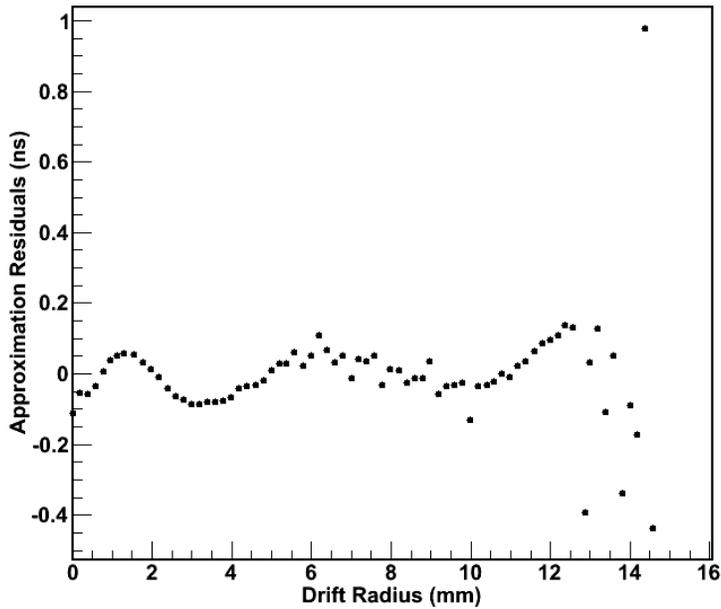

**Figure 13: Residuals of GARFIELD generated RT-function at 298K and GARFIELD generated RT-function at 287 K with the addition of the temperature correction function (12) scaled to temperature correction function (Figure 12) scaled to the same temperature difference of 12 K.**

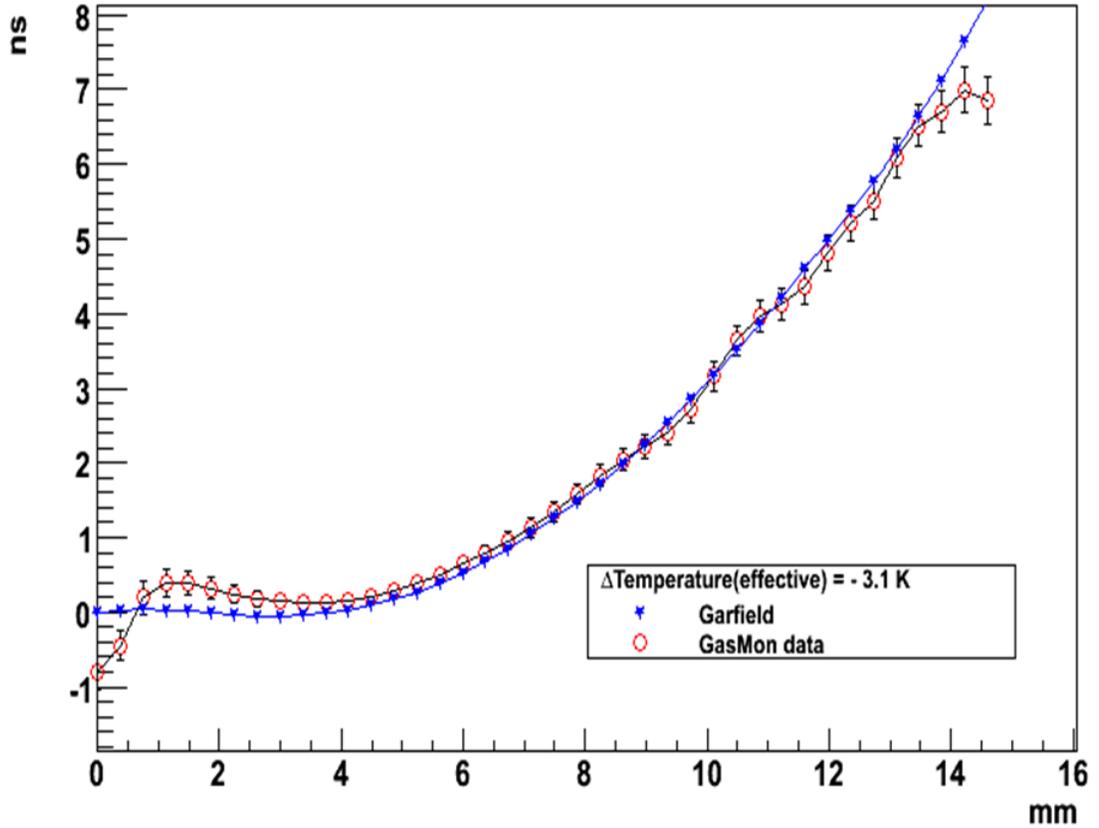

**Figure 14: Residuals of two GMC RT-functions with a temperature difference of -3.1 K and temperature correction function (Figure 12) scaled to the same temperature difference.**

### 2.3.3 Magnetic Field Corrections

The ATLAS Muon Spectrometer Magnet System comprises a set of eight air-core superconducting coils of 25 m length, which provide a toroidal magnetic field in the barrel region. The average field strength is ~0.5 T in the barrel region, but can vary to ~1 T. Typically the field direction is parallel to the monitored drift tube wires with up to ~0.4 T variance in this direction. However, significant components can be found along the muon tracks.

#### 2.3.3.1 Magnetic Field Effects on Drift Times

External magnetic fields cause drift electrons to deviate from their radial paths into curved trajectories. This causes an elongation in the drift times. The drift electrons' trajectory can be described by Newton's Law with a nearly linear damping force and a Lorentz force (Eq. 2)

(2) $$\ddot{x} = -(\frac{\dot{x}}{\tau})^{1+\varepsilon} + \frac{q}{m}[\mathbf{E}+\dot{\mathbf{x}} \times \mathbf{B}]$$

This is a class of Langevin equation, which describes biased Brownian motion [11]. The drag force parameter, $\varepsilon$, has the experimentally determined value of ~0.007, and $\tau$ is the mean free time between collisions. Approximate solutions to these equations are well known. The approximate perturbation solution is given in Eq. 3 [11].

(3)
$$t(r,B) = t(r, B=0) + B^{2-\varepsilon} * \int_{r_{wire}}^{r} \frac{v_{B=0}^{1+\varepsilon}}{E^{2-\varepsilon}} dr$$

This solution has two important features: It is both separable and factorable. The first term is the drift time as a function of drift radius when there is no external magnetic field. The second is the perturbation correction to the drift time which is an integral of the drift velocity and the radial electric field, multiplied by a factor of $B^{2-\varepsilon}$. In this notation the magnetic field magnitude, B, is the norm of the components of the magnetic field perpendicular to the drift electrons' paths: the component parallel to the muon track, and parallel to the MDT anode wire. We define this as the magnitude of the effective magnetic field, $B_{eff}$. The magnetic field component parallel to the drift electrons' paths can also elongate drift times, but the effects are second order. The perturbation-integral depends on the radial coordinate and local geometry. It is a constant with respect to the magnetic field. Therefore, the overall correction is a function of the radius scaled by a nonlinear B-field term. As a result, for small changes in the magnetic field we expect the changes in drift time to scale nearly quadratically with the effective magnetic field [9]

### 2.3.3.2 GARFIELD Simulations of Non-Zero Magnetic Fields

The effect of external magnetic fields was verified using GARFIELD. The simulations were similar to those made for temperature and pressure and follow the general procedure explained in section II.3.A . However, the variations are of a vector, not scalar field as before. As noted above, the first order effects on drift times depend on the magnetic field components perpendicular to the drift electrons' paths, and thus it is sufficient to treat variations in the magnetic field along the anode wire alone. However, the correction function (section 3, C) was tested on three component magnetic fields for completeness. A plot of $dT_{max}^{RT}$ versus $B_{eff}$ is shown in Figure 15, in which the quadratic dependence of the changes in drift times with magnetic field is verified. $T_{max}^{RT}$ increases by ~ 22 ns for $B_{eff}$ ~0.5 T, or 3% increase. This corresponds to ~450 μm degradation of the nominal MDT resolution if the magnetic field is unaccounted for, which is well above the error budget of ~100 μm.

### 2.3.3.3 GARFIELD Calibrated Magnetic Field Correction Function

The calibrated correction function for external magnetic fields was computed in much the same way as the temperature and pressure correction functions: An ensemble of RTs are generated and averaged for each GARFIELD magnetic field simulation, from 0.05T to 0.5T in increments of 0.05T. The RTs are mapped into Sobolev space and fit with Chebyshev polynomials. Iterative difference functions, dTR curves, are computed for every possible magnetic field combination. The dTR curves are normalized to the same base-line correction $dTR(B_{eff} = .05\ T)$ by convention. A set of dTR functions computed for the zero B-Field baseline can be seen in Figure 16. The normalization procedure is slightly more complicated in this case due to the inherent nonlinearity of the drift time corrections. However, the dTR curves are simply the differences between the drift time functions, t(r,B), shown in (3). For a particular pair of B-Fields, $B_1$ and $B_2$, the drift time difference function, $dTR(B_1, B_2)$, has the following form (Eq. 4), where I(r) is the integral of the electric field and drift velocity and is the same for all such dTR curves.

(4)
$$dTR(B_1, B_2) = (B_2^{2-\varepsilon} - B_1^{2-\varepsilon})*I(r)$$

The B-Field dependence is a simple difference of squares factor and is multiplied by a radially-dependent function. Therefore, the drift time corrections can be normalized to the same correction by the appropriate scaling with a ratio of two differences of squares terms. To first order, the dTR curves have the same shape but are scaled by different multiplicative factors. For example, dTR(0.2, 0.5 T) is equal to dTR(0, 0.05 T) multiplied by ($10^{2-\varepsilon}$- $4^{2-\varepsilon}$). The rest of the algorithm is very similar to the temperature case. Each dTR bin has 55 points which are averaged, filtered with a 2*sigma cut, and then re-averaged. The remaining calibration curve represents a scalable, radius-bin-by-bin, drift time correction function that can be added to a 293 K : 3 bar : 0 T gas monitor RT functions to give a drift time-drift radius calibration that is compensated for gas composition and magnetic field (Figure 17). More generally, the magnetic field correction function can be added to a temperature and pressure compensated RT function- all corrections are independent to first order.

Similar to the temperature correction procedure, an internal consistency check of the algorithm with GARFIELD can be made by adding the properly normalized calibration function to a baseline RT, and measuring how well it approximates a simulated RT function with a strong magnetic field. Moreover it can be checked whether the assumption that, to first order, drift times only depend on the magnitude of the B-Field that is perpendicular to the drift electrons' paths by testing the correction function on a 3-component B-Field. Figure 18 shows the residuals between a simulated RT function with the magnetic field (.3 T, .3 T, .3 T), and an RT formed by the addition of an RT function simulated with zero B-Field and the calibration function normalized to an effective field of $\sqrt{2}$*0.3 T (the two perpendicular components added in quadrature). It can be seen that the calibration curve underestimates the drift time changes. However, the residuals are everywhere less than ~0.5 ns with drift time corrections two orders of magnitude larger. The induced resolution degradation in this case is only ~10 μm.

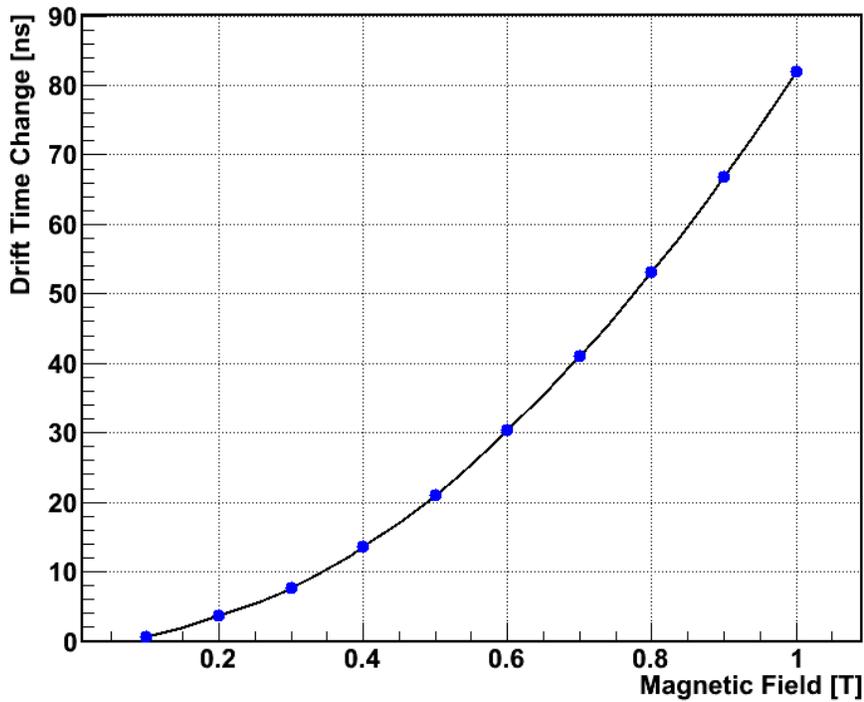

Figure 15: Change in maximum drift time of GARFIELD generated RT-functions versus the effective magnetic field (Equation 3)

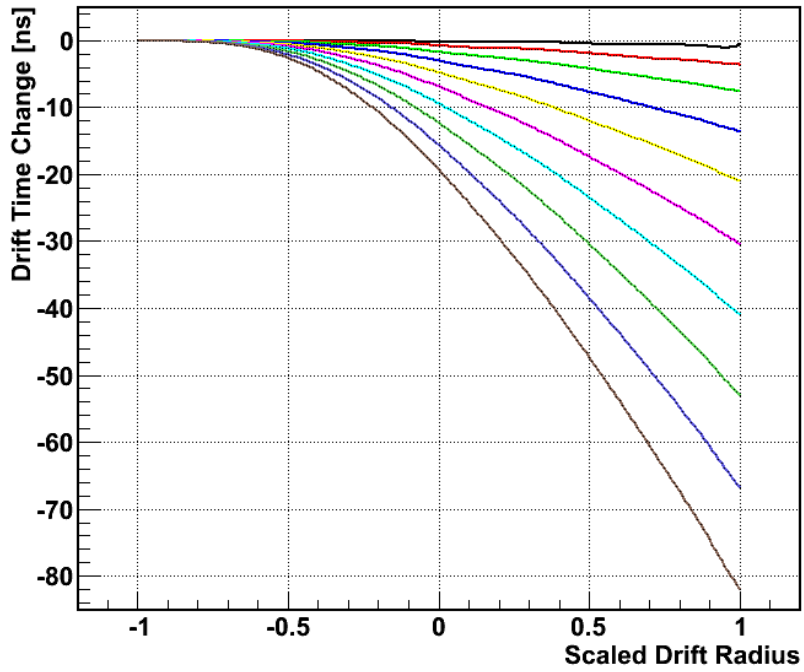

**Figure 16:** Drift time differences between GARFIELD RT function with no magnetic field and increasing magnetic fields (steps of 0.05 T)

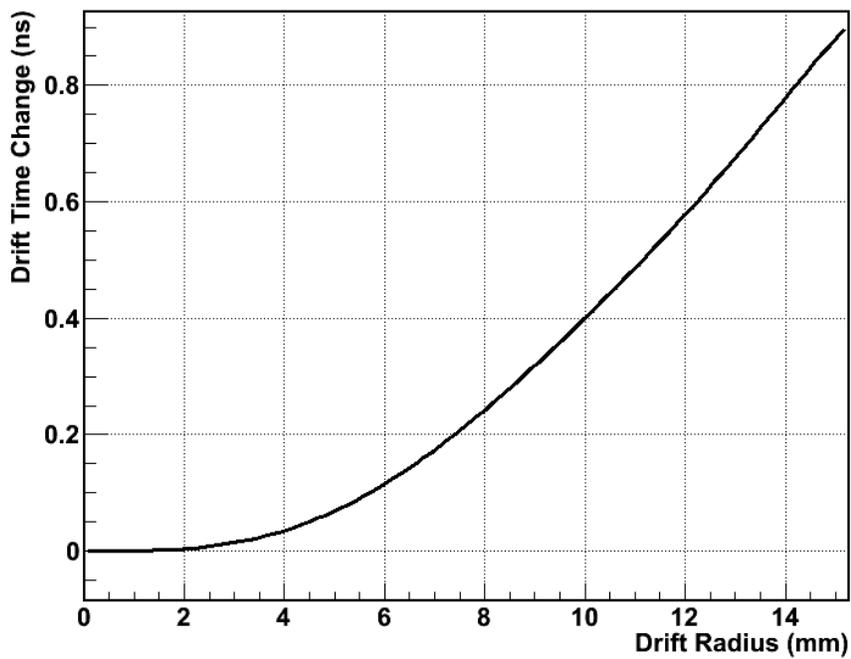

**Figure 17:** Calibrated correction function normalized to an effective magnetic field of 0.05T.

### 2.3.4 Chamber Averaged Effective Magnetic Fields

The general gas monitor correction algorithm (II.5) assigns an RT function for every chamber, constructed from the gas monitor RT functions and the addition of the various corrections functions scaled to local chamber field conditions. Average chamber temperature readings are well known. However, it is much more difficult to assign magnetic field averages over chambers, particularly when the field varies significantly. A field map of the EIL4 chambers, shown in Figure 19, illustrates this point. The magnitude of the field can be seen to vary from ~0 T to 0.9 T. This variation incurs maximum drift times to change over a range of nearly 70 ns.

A simple measure of the average effective B-field in a chamber is a volume integral of the scalar field $B_{eff}$ divided by the integration volume. $B_{eff}$ has no directionality, though it depends on the components of the B-Field. If the differential volume elements are constant, a simple sigma and mean for $B_{eff}$ can be defined. The coordinates and sizes of the MDT chambers can be extracted from the AMDB, the ATLAS Muon Spectrometer geometric database. The approximate magnetic field map for each chamber type can be computed from the most recent spectrometer magnetic field map and a Biot-Savart interpolation procedure.

The B-Field is given in terms of the global cylindrical coordinates (r, ρ, z). However as mentioned, it is more useful to project the vector field onto another orthogonal coordinate system with basis: the anode wire, the muon track, and the drift electrons' path of closest approach. The anode wires in all MDT chambers are approximately in the ϕ direction; they curl around the z-axis of symmetry. Furthermore, the muons from collisions travel in approximately radial paths from the interaction point-radial in the spherical sense, ρ. It should be noted that the latter approximation will not be true for cosmic rays. With these assumptions the effective magnetic field can be computed with Eq. 5.

$$B_{eff} = \sqrt{B_\phi^2 + B_r^2\left(\frac{r}{\rho}\right)^2 + 2B_rB_z\left(\frac{rz}{\rho^2}\right) + B_z^2\left(\frac{z}{\rho}\right)^2} \qquad (5)$$

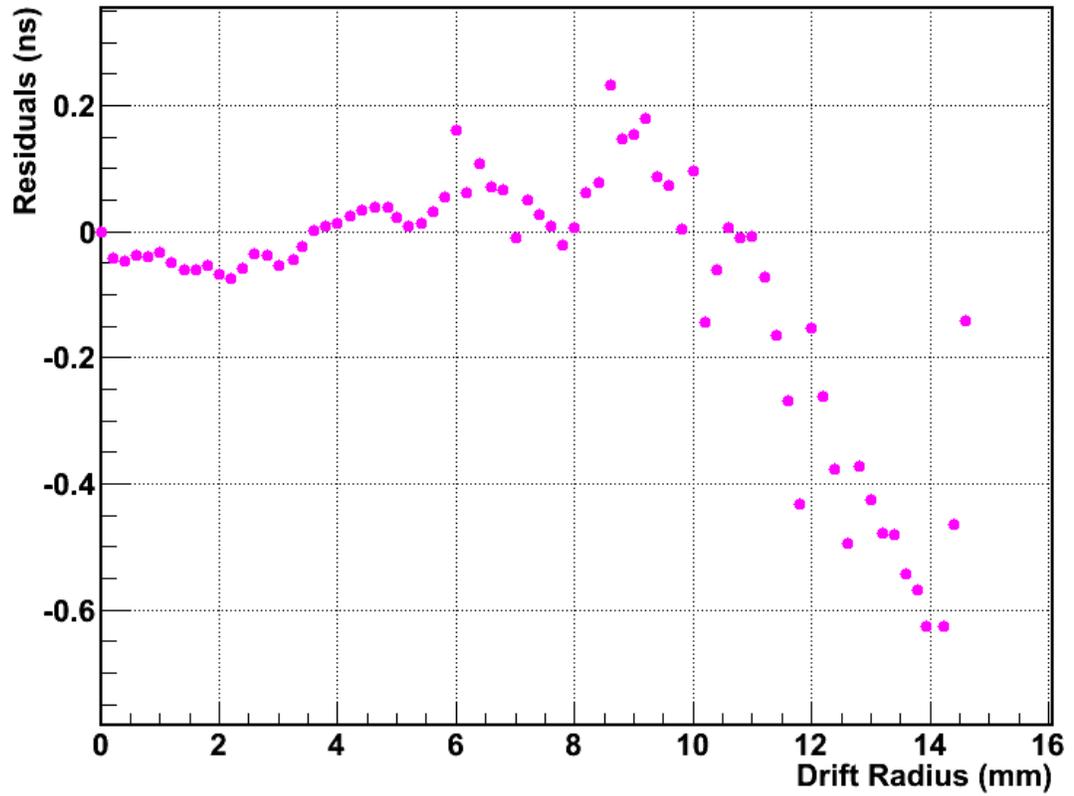

**Figure 18:** Residuals of GARFIELD RT-function with 3-component B-Field (.3T, .3T, .3T) and RT function formed from the addition of GARFIELD RT function with zero B-Field and the scaled correction function (18).

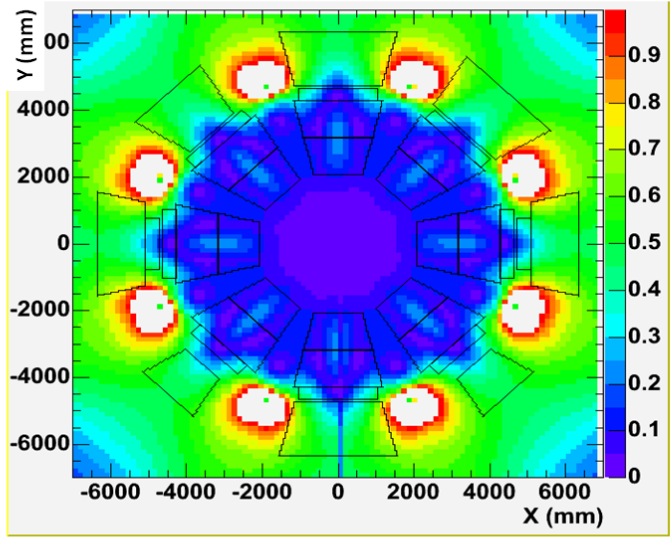

**Figure 19: Map of effective B-Field in ATLAS Muon Spectrometer inner endcap station. Trapezoids indicate chamber placement. Only higher radius chambers have significant magnetic fields.**

Assigning and average effective magnetic field to regions in space where the field is not uniform will induce some error in the drift time corrections, and consequently the position measurements. Therefore, any gradient in $B_{eff}$ induces a characteristic increase in the MDT resolution. If a magnetic field correction procedure is to be added to the gas monitor calibration algorithm, a measure is needed to assess the effect of neglecting gradients. We define an error metric per chamber, the average systematic MDT resolution degradation, or ASRD, to be the resolution that must be added in quadrature to the nominal tube resolution on average for a given chamber. The chamber ASRD is computed in the following way:

1. Divide every chamber into regular volume elements.
2. Scan through each volume element in a given chamber iteratively.
3. At each volume element compute the bin-by-bin drift time corrections from the average effective field, $B_{avg}^{eff}$, and the actual effective field, $B_{vol}^{eff}$.
4. Take the difference in the drift time corrections at each bin.
5. Average over all bins and multiply by the average electron drift velocity. This number represents the resolution degradation of a tube in the given volume element.
6. Repeat for all volume elements and average over all the local resolution degradations. This number is the given chamber ASRD.
7. Repeat for all chambers.

A histogram of the ASRD for all sections of the barrel and endcap is shown in Figure 20. The ASRD for most endcap chambers is negligible populates the first bin.

In order to have the necessary muon momentum resolution we require MDT resolution of ≲ 100 μm. The nominal tube resolution is ~80 μm without the magnetic field correction procedure. Therefore, the ASRD for a given chamber should not exceed ~50 μm. Under this criterion, only 70% of the chambers can be corrected for magnetic field in this way. However, this includes nearly 90% of the endcap chambers with the several exceptions. For each wheel: 90% in endcap inner wheel (EI),

100% in endcap middle wheel (EM), 100% in endcap outer wheel (EO), 10% in endcap extra (EE), 54% in barrel inner, 63% in barrel middle (BM), and 55% in barrel outer (BO). This implies that most endcap chambers can be corrected for magnetic field without exceeding the error budget. However, corrections in the barrel must be made at the hit-level.

In general, magnetic field corrections are harder to make than temperature and pressure corrections within the resolution specifications. This difficulty is function of field gradients and the inherent non-linearity of the correction. Resolution will be degraded by this procedure through a coupling of these effects. For example, if a strong field is coupled with a modest gradient, the resolution can still be highly degraded. The opposite case is also true. A plot of the ASRD for each chamber versus the product of $B_{eff}^{avg}$ and $\sigma_{eff}^{B}$, defined as the field convolution, is shown in Figure 21. From this plot we can see the strong positive correlation between resolution degradation and field convolution.

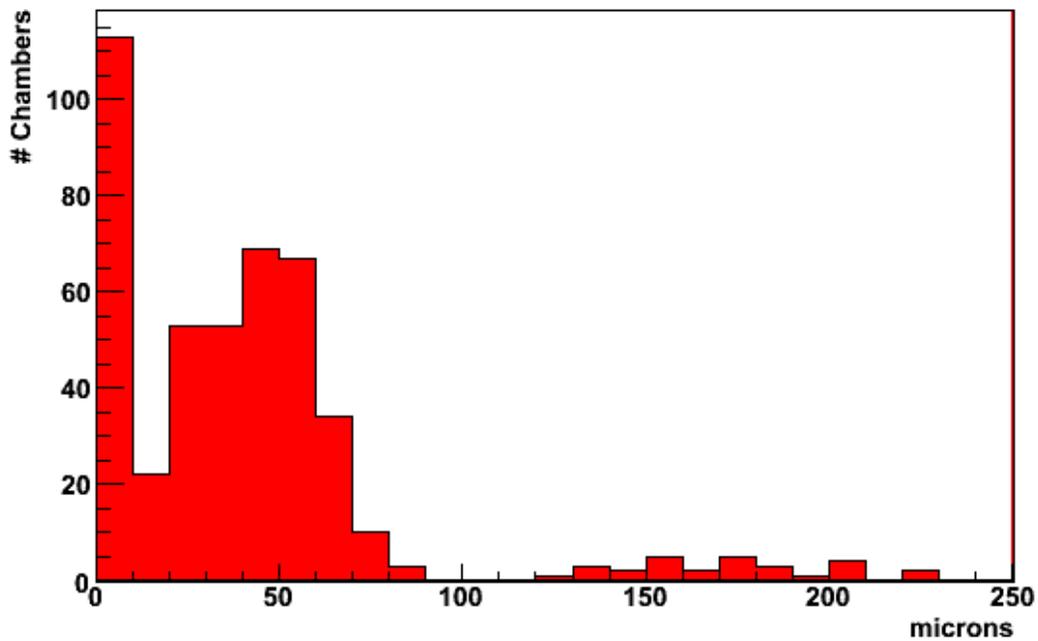

Figure 20: Average systematic resolution degradation for the inner region of the endcap (II.3.D)

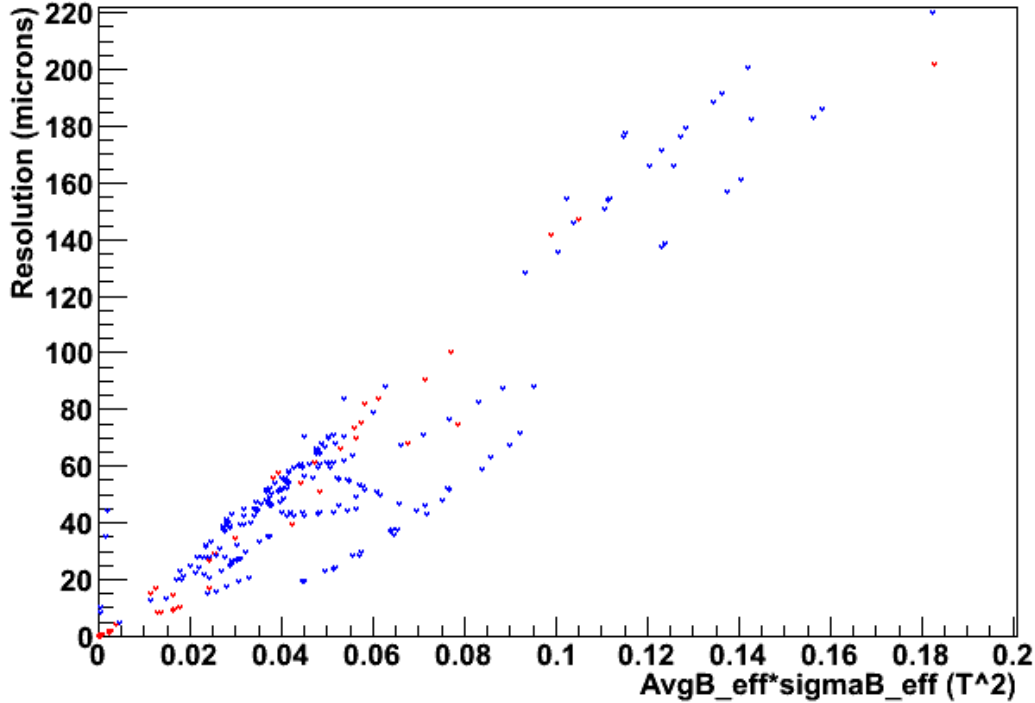

**Figure 21: Resolution degradation versus the product of the average B field and its variation across a single chamber.**

## 2.4 Automated Gas Monitor Calibrations

For the LHC startup phase, hit rates are to be modest. Under these circumstances, the MDT electric fields are nearly static, and hit density alone does not appreciably incur large drift time changes. In this regime, GARFIELD-simulation corrections to gas monitor RT functions provide verifiably accurate, easily applicable, MDT calibrations. Moreover, the drift time correction procedures merely supplement the normal tasks of the gas monitor system. The ease with which large scale calibrations, for the entire MDT system, can be made in this way is a function of the simplicity of the drift time corrections: once the normalized correction functions are constructed, they can simply be scaled to represent local environments. One complication is choosing regions over which the spectrometer can be assumed to have uniform field values, such that a set of average conditions accurately codifies the region. We choose the spectrometer's natural unit, the MDT chamber for this task. To apply the gas monitor calibration algorithm in principle, we require every chamber's time and space averaged temperature, pressure, and effective magnetic field. In practice, however, the magnetic field is virtually static and most chambers' pressures are systematically shifted ~5mbar above 3bar. Daily-average chamber temperatures are well known.

### 2.4.1 Daily Procedure

The gas monitor system produces RT functions every two hours from an auto-calibration procedure (Gas Monitor RT Function Generation section II.2.C). The twelve RT functions generated daily are averaged into a universal RT function, URT [13]. Typically, MDT calibration constants are required with 24-hour latency. The URT represents a given day's baseline calibration for the MDT system gas, and can be corrected for local chamber conditions via the addition of properly scaled correction functions. The daily output is thus a set of RT functions for every chamber, which for each chamber represents the local calibration for gas composition, temperature, pressure, and where applicable, magnetic field. The virtue of this method is that it can be automated, if coupled with inherent security mechanisms.

Each chamber's temperature is measured daily as an average of several onboard sensor measurements. If sensors appear to be broken, the output readings are unphysical – the chamber's temperature is replaced by the interpolation of the cavern's vertical temperature profile, if available, or the cavern's average temperature. The temperature is read-in for each chamber, the correction function is scaled to that temperature, and then added bin-by-bin to the URT. The characteristic pressure corrections to first approximation are systematic shifts ~5 mbar above 3 bar. Therefore, the pressure correction function is scaled to 5mbar and added bin-by-bin to each chamber's temperature compensated RT function. The temperature and pressure compensated RT functions for every chamber are sent to the ATLAS calibration centers to be used by the MDT reconstruction software.

Chamber-level magnetic field compensation incurs more error than the other field corrections. This error is due to sizeable gradients and the inherent nonlinearity of the drift time changes with effective magnetic field strength (Chamber Averaged Effective Magnetic Fields II.3.D). For this reason, URT calibrations for pressure, temperature, and magnetic field are kept separate from the simpler pressure and temperature cases. The magnetic field corrections are made by computing the average effective field for each chamber from the most recent field map, scaling the correction functions for each chamber value, and adding the scaled correction functions bin-by-bin to the temperature/pressure corrected RT function for each chamber. These temperature, pressure, and magnetic field compensated RT functions for every chamber are sent to the ATLAS calibration centers to be used by the MDT reconstruction software.

The automated gas monitor system calibration procedure has been validated on ATLAS preliminary runs (section II.7) [10]. Track segment reconstruction with cosmic rays was evaluated via the hit residual distributions from an ensemble of chambers. The residual widths were found to be near 100μm, which is close to design expectations [10].

## 2.5 Space Charge Effects

Space charge effects occur when ions from electron avalanches distort the MDT electric fields. Because the ionization process follows a Poisson distribution, we expect the resolution degradation to be approximately proportional to the square root of the background rate [12]. However, these effects are negligible at LHC initial running where luminosity is not expected to be high. Moreover, it has been shown that the systematic increase in the resolution can be less than 30 μm given a parameterized correction function similar to those discussed in this paper, even at the maximum expected rate of 1500 Hz/cm [12]. In practice, the event rates can be constructed at the chamber level from monitored chamber currents, from which, space charge effects could be accounted for.

## 2.6 Application of Calibration Program in ATLAS Cosmic Ray Commissioning Data

An important test of the Gas Monitor based calibration program was instituted on a cosmic ray run from September 2008 [13]. In the test, the quality of track segment reconstruction on an ensemble of MDT chambers was assessed using the residuals metric [8]. The residual metric is defined as the difference between the impact parameter for a given hit MDT and the radial distance of the best-fit track when that MDT is excluded from the fit. This metric stands proxy for the MDT resolution. 951 chambers with more than 2000 segment hits were fit with double Gaussians, with the width and mean extracted from each. It was found that the average of the means was very close to zero and within the allowable error budget. Moreover the peak of the residual widths was ~107 μm with 90% of the chambers having widths of less than 140μm. With the compensation for test uncertainties, the MDT resolution was ~100 μm and the residual means were near zero [13]. The test showed that the resolution from this method approached the design specifications.

## 3  Conclusion

The Gas Monitor Chamber Based Calibration program was shown to offer a computationally simple approach to the MDT calibration task as a natural extension of normal gas monitoring. The program provides RT function calibration constants daily for all MDT chambers. The chamber specific RT functions represent local field compensations added to the Universal Gas Monitor RT functions. The local drift time corrections are determined from daily averaged chamber temperatures, magnetic field maps, and GARFIELD calibrated drift time correction curves. Track quality assessment from residual distributions of cosmic ray runs, have shown this program to be suitable for the startup phase of the LHC. For luminosities approaching the full design expectations of the LHC, track reconstruction will be done exclusively at the hit level.